\documentclass[fleqn,usenatbib]{mnras}

\usepackage{newtxtext,newtxmath}

\usepackage[T1]{fontenc}
\usepackage{underscore}
\usepackage{array}
\usepackage{multirow}
\DeclareRobustCommand{\VAN}[3]{#2}
\let\VANthebibliography\thebibliography
\def\thebibliography{\DeclareRobustCommand{\VAN}[3]{##3}\VANthebibliography}

\usepackage{graphicx}	
\usepackage{amsmath}	

\newcommand{\coralie}{{CORALIE}}

\renewcommand{\arraystretch}{1.25}
\pdfoutput=1

\usepackage{xcolor}
\usepackage{soul}

\title[EBLM XVI. Alignment of EBLM J0021-16]{The EBLM project XVI. Moderate spin-orbit misalignment of the low mass eclipsing binary EBLM J0021-16}

\author[Becca Spejcher et al.]{
Becca Spejcher$^{1*}$$^{\href{https://orcid.org/0009-0003-7036-2840}{\includegraphics[scale=0.5]{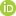}}}$,\author[David V. Martin]{
Becca Spejcher$^{1*}$$^{\href{https://orcid.org/0000-0002-7595-6360}{\includegraphics[scale=0.5]{orcid.jpg}}}$,
\\
*Rebecca.Spejcher@tufts.edu \\
Authors to add (grouped, no particular order):\\
AST-192 crew: David, Max, Jake, Andy, Wata \\
EBLM David collaborators: Alison, Shelby, Ritika, Noah V\\
EBLM broader collaborators: Amaury, Pierre, Adrian Barker,...\\
Geneva collaborators: Stephane, François, TBD...\\
Affiliations are listed at the end of the paper.
}}

\author[Spejcher et al.]{
        Becca Spejcher$^{1*}$$^{\href{https://orcid.org/0009-0003-7036-2840}{\includegraphics[scale=0.5]{orcid.jpg}}}$,
        David V. Martin$^{1}$$^{\href{https://orcid.org/0000-0002-7595-6360}{\includegraphics[scale=0.5]{orcid.jpg}}}$,
        Jake Pandina$^{1}$$^{\href{https://orcid.org/0009-0005-1978-0320}{\includegraphics[scale=0.5]{orcid.jpg}}}$,
        Andy Zhang$^{1}$$^{\href{https://orcid.org/0000-0000-0000-0000}{\includegraphics[scale=0.5]{orcid.jpg}}}$,
        Max Ammons$^{1}$$^{\href{https://orcid.org/0009-0002-9966-5565}{\includegraphics[scale=0.5]{orcid.jpg}}}$,\newauthor
        Wata Tubthong$^{1}$$^{\href{https://orcid.org/0000-0002-7907-2634}{\includegraphics[scale=0.5]{orcid.jpg}}}$,
        Amaury Triaud$^{2}$$^{\href{https://orcid.org/0000-0002-5510-8751}{\includegraphics[scale=0.5]{orcid.jpg}}}$,
        Ritika Sethi$^{3}$$^{\href{https://orcid.org/0000-0002-6576-3346}{\includegraphics[scale=0.5]{orcid.jpg}}}$,
        Noah Vowell$^{4}$$^{\href{https://orcid.org/0000-0002-0701-4005}{\includegraphics[scale=0.5]{orcid.jpg}}}$,
        Adrian Barker$^{5}$$^{\href{https://orcid.org/0000-0003-4397-7332}{\includegraphics[scale=0.5]{orcid.jpg}}}$,\newauthor
        Pierre Maxted$^{6}$$^{\href{https://orcid.org/0000-0003-3794-1317}{\includegraphics[scale=0.5]{orcid.jpg}}}$,
        Alison Duck$^{8,7}$$^{\href{https://orcid.org/0000-0002-4531-6899}{\includegraphics[scale=0.5]{orcid.jpg}}}$,
        Shelby Summers$^{7}$$^{\href{https://orcid.org/0009-0006-8371-7227}{\includegraphics[scale=0.5]{orcid.jpg}}}$,
        François Bouchy$^{9}$$^{\href{https://orcid.org/0000-0002-7613-393X}{\includegraphics[scale=0.5]{orcid.jpg}}}$,
        Monika Lendl$^{9}$$^{\href{https://orcid.org/0000-0001-9699-1459}{\includegraphics[scale=0.5]{orcid.jpg}}}$,\newauthor
        Maxime Marmier$^{9}$$^{\href{https://orcid.org/0000-0001-5630-1396}{\includegraphics[scale=0.5]{orcid.jpg}}}$,     
        Malte Tewes$^{10}$$^{\href{https://orcid.org/0000-0002-1155-8689}{\includegraphics[scale=0.5]{orcid.jpg}}}$,
        Stéphane Udry$^{9}$$^{\href{https://orcid.org/0000-0001-7576-6236}{\includegraphics[scale=0.5]{orcid.jpg}}}$\\
$^{1}$Department of Physics and Astronomy, Tufts University, Medford, MA 02155, USA\\
$^{2}$School of Physics and Astronomy, University of Birmingham, Edgbaston, Birmingham B15 2TT, UK\\
$^{3}$Department of Physics, Massachusetts Institute of Technology, Cambridge, MA 02139, USA\\
$^{4}$Department of Physics and Astronomy, Michigan State University, East Lansing, MI 48824, USA\\
$^{5}$School of Mathematics, University of Leeds, Leeds LS2 9JT, UK \\
$^{6}$Astrophysics Group, Keele University, Keele, Staffordshire, ST5 5BG, UK\\
$^{7}$Department of Astronomy, The Ohio State University, Columbus, OH 43210, USA\\
$^{8}$Jet Propulsion Laboratory, California Institute of Technology, 4800 Oak Grove Drive, Pasadena, CA 91109, USA\\
$^{9}$Observatoire Astronomique de l’Université de Genève, Chemin Pegasi 51, CH-1290 Versoix, Switzerland\\
$^{10}$ Universität Bonn, Argelander-Institut für Astronomie, Auf dem Hügel 71, 53121 Bonn, Germany
}

\date{MNRAS Submission September 25 2025}

\pubyear{2025}

\begin{document}
\label{firstpage}
\pagerange{\pageref{firstpage}--\pageref{lastpage}}
\maketitle

\begin{abstract}

Thousands of tight ($<1$ AU) main sequence binaries have been discovered, but it is uncertain how they formed. There is likely too much angular momentum in a collapsing, fragmenting protostellar cloud to form such binaries in situ, suggesting some post processing. One probe of a binary's dynamical history is the angle between the stellar spin and orbital axes -- its obliquity. The classical method for determining stellar obliquity is the Rossiter-McLaughlin effect. It has been applied to over 100 hot Jupiters, but less than a dozen stellar binaries. In this paper, we present the Rossiter-McLaughlin measurement of EBLM J0021-16, a $0.19M_\odot$ M-dwarf eclipsing a $1.05M_\odot$ G-dwarf on a 5.97 day, almost-circular orbit. We combine CORALIE spectroscopy with TESS photometry and a measured primary star rotation period of 7.04 days, according to star spot modulation. We show that the orbital axis is misaligned with the primary star's spin axis, with a true 3D obliquity of $\psi=28.9\pm2.1^{\circ}$. EBLM J0021-16, being neither spin-orbit aligned nor synchronized, yet with an almost circular orbit, is a curious case for tidal evolution in tight binaries. It becomes one of a handful of eclipsing binaries with true obliquity measurements. Finally, we derive the M-dwarf's mass and radius to a fractional precision better than 1\%. The radius of the M-dwarf is inflated by 6\% ($7.4\sigma$) with respect to stellar models, consistent with many other M-dwarfs in the literature.

\end{abstract}

\begin{keywords}
stars: binaries-eclipsing, low-mass, fundamental parameters; techniques: photometric, spectroscopic
\end{keywords}



\section{Introduction}

The formation and evolution of hot Jupiters are still not well-understood. One indicator of their origin is the obliquity between the planet's orbit and the star's rotation. This can be measured through the Rossiter-McLaughlin (RM) effect, where spectroscopy is taken during the planet's transit \citep{Rossiter1924,McLaughlin1924,Queloz2000b,Triaud2017}. The RM effect has been observed for over 100 
hot Jupiters, 
revealing diverse orbits: coplanar, misaligned, polar, and even retrograde. A misalignment might speak to a dynamical history (e.g. scattering or Kozai-Lidov cycles), whereas an alignment might imply a more placid history (e.g. in situ formation or smooth disc migration). To complicate the matter, star-planet tidal interactions tend to re-align the planet's orbit over time, potentially washing away the original formation signature \citep[e.g.][]{BO2009}. Overall, whilst the origin of hot Jupiters is yet to be settled, measurements of spin-orbit alignment are an integral piece of the puzzle \citep{Winn2005,Fabrycky2007,Triaud2010,Albrecht2012,Dawson2018,Rice2022}.

Another type of astrophysical object with an equally perplexing origin is a tight stellar binary. Collapsing and fragmenting protostellar clouds have too much angular momentum to place two stars on a separation of a fraction of an AU (orbital periods of $\sim$days, \citealt{Tohline2002,Moe2018}). Nevertheless, thousands of short-period main sequence binaries have been discovered \citep{Raghavan2010,Prsa2011,Jayasinghe2018,Prsa2022,Mowlavi2023}. Whilst tight binaries and hot Jupiters both have a complex and debated origin, tight binaries have not received the same level of observational attention; only 6 sky-projected obliquities have been measured for main sequence eclipsing binaries, compared to 100+ for hot Jupiters \citep{Marcussen2022}.

In 2010, we created the EBLM (Eclipsing Binary Low Mass) survey. The sample consists of M-dwarfs ($\approx 0.08 - 0.6 M_\odot$) eclipsing more massive F/G/K dwarfs. These systems were discovered as ``false positives'' from the WASP hot Jupiter survey. Giant planets, brown dwarfs, and the lowest mass M-dwarfs all have a similar radius ($\sim 1R_{\rm Jup}$) despite spanning two orders of magnitude in mass. This means that the photometric signature of an M-dwarf eclipsing an F/G/K dwarf is similar to that of a transiting hot Jupiter, i.e. a $\sim 0.5-5\%$ dip. There are multiple goals for the EBLM project, one of which is increasing the number of precise obliquities of eclipsing binaries (see the review in \citealt{Maxted2023}). So far, the EBLM project has published obliquities for WASP-30 (an eclipsing brown dwarf) \& EBLM J1219-39 \citep{Triaud2013}, EBLM J0218-31 \citep{Gill2019}, and the circumbinary planet host EBLM J0608-59 (also known as TOI-1338, BEBOP-1) \citep{hodzic2020}. 

In this paper, we present the latest target in the EBLM series of RM measurements: EBLM J0021-16. The system consists of a $0.19M_\odot$ M-dwarf eclipsing a $1.05M_\odot$ primary star in a 6-day orbit. The radial velocities for both the RM effect and Keplerian orbit were observed with the CORALIE spectrograph. TESS photometry reveals primary eclipses, secondary eclipses, and a starspot-induced rotation signature of the primary star. The organization of this paper is as follows. In Sect.~\ref{section:observations} we describe the spectroscopic and photometric observations. In Sect.~\ref{section:methods} we outline the data-fitting methods. We show the results in Sect. ~\ref{section:results} and then in Sect.~\ref{section:discussion} we discuss implications with respect to tidal physics, other measured obliquities in the literature, and M-dwarf radius inflation, before concluding.


\section{Observations}\label{section:observations}

\begin{table}
\caption{Summary of observational properties.}              
\label{Observation_table}      
\centering                                    
\begin{tabular}{c l l  l l l l l l l l l l l l l }  

\hline\hline                        
 
 & EBLM J0021$-$16  \\
 
\hline 
 TIC
 & 406352710 \\
 SAO 
 & 147227 \\
 GSC
 & 05839-00998\\
 2MASS 
 & 00210077-1607285\\
 Gaia ID
 & Gaia DR3 2368434190390080000\\\\
 
RA ($\alpha$) [BCRS]
& $00^{\rm h}21'00.77''$ \\
& $5.253074^{\circ}$ \\

Dec ($\delta$) [BCRS]
& $-16^{\circ}07' 28.61"$\\
& $-16.124906^{\circ}$\\ \\

CORALIE obs \#
& 67\\

CORALIE obs $\Delta$ t [days]
& 1950 \\

{\it TESS} Sectors
 
& \#29, \#107 (Coming August 2026)\\ 

$TESS$-mag
& $9.18744 \pm 0.0061$ \\

$B$-mag
& $10.49 \pm 0.04$ \\

$G$-mag
& $9.631901 \pm 0.002793$\\

$V$-mag
& $9.80 \pm 0.03$ \\

$G_{\rm BP} - G_{\rm RP}$
& 	
0.826 \\

Parallax [mas]
& $6.6914 \pm 0.0206$ \\ 
Distance [pc]
& 148.1527 \\ 
RUWE
& 1.1498165 \\

\hline

\end{tabular}
\end{table} 

\subsection{CORALIE Radial Velocity Spectroscopy}

\begin{figure}
    \includegraphics[width=0.49\textwidth,trim={6.2cm 6.9cm 15cm 8cm},clip]{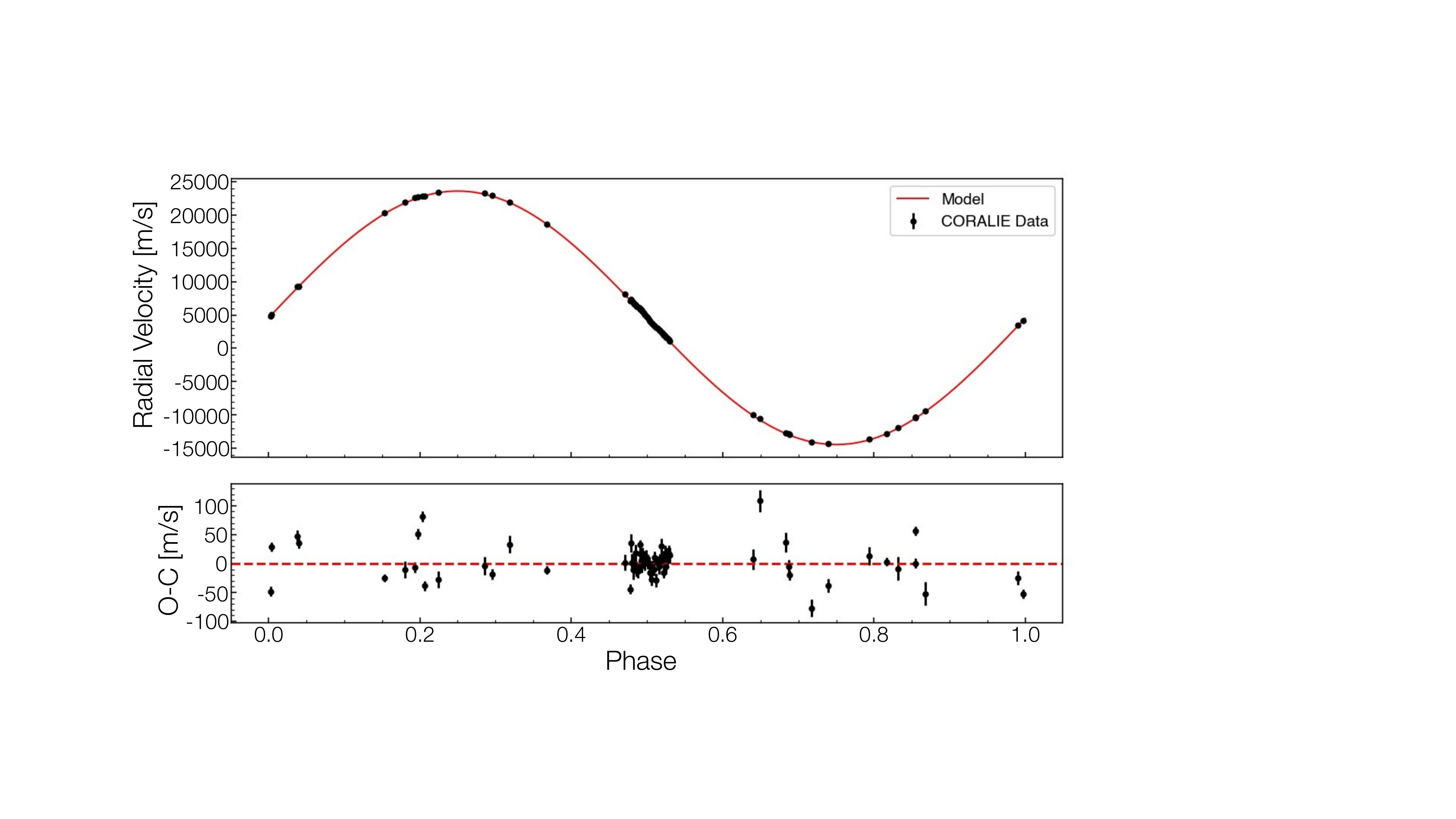} 
    \caption{Radial velocity observations of EBLM J0021-16, phase-folded on the best fitting orbital period. The CORALIE data are shown by the black dots, with error bars on the order of m/s and are invisible at this scale. The red line shows the {\tt \textsc{PyMC3}} best fit model \textit{with} the RM fit as well. The RM here is not visible because it happens on the order of m/s, which cannot be seen in the Keplerian, which is on the order of km/s.}\label{fig:RVs}
\end{figure}

\begin{figure}
    \includegraphics[width=0.49\textwidth]{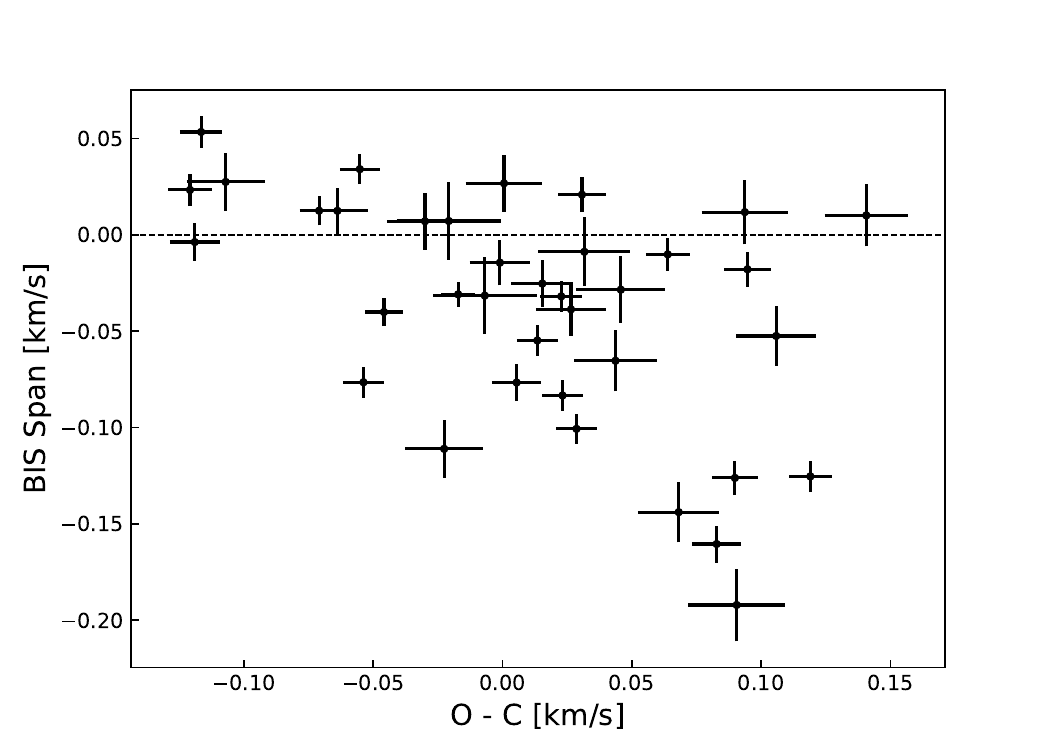}
    \caption{Relationship between the CCF bisector span and RV residuals from a Keplerian + linear drift model. The RM data are excluded. As noted in  \citet{Triaud2017}, there is clear anti-correlation between the RV residuals and the bisector. This implies the presence of stellar activity, which is also seen in the star spot modulation in Fig. \ref{fig:PDCSAP vs SAP Transit}. }\label{fig: Bisector Plot}
\end{figure}

\begin{figure*}
    \includegraphics[width=0.99\textwidth]{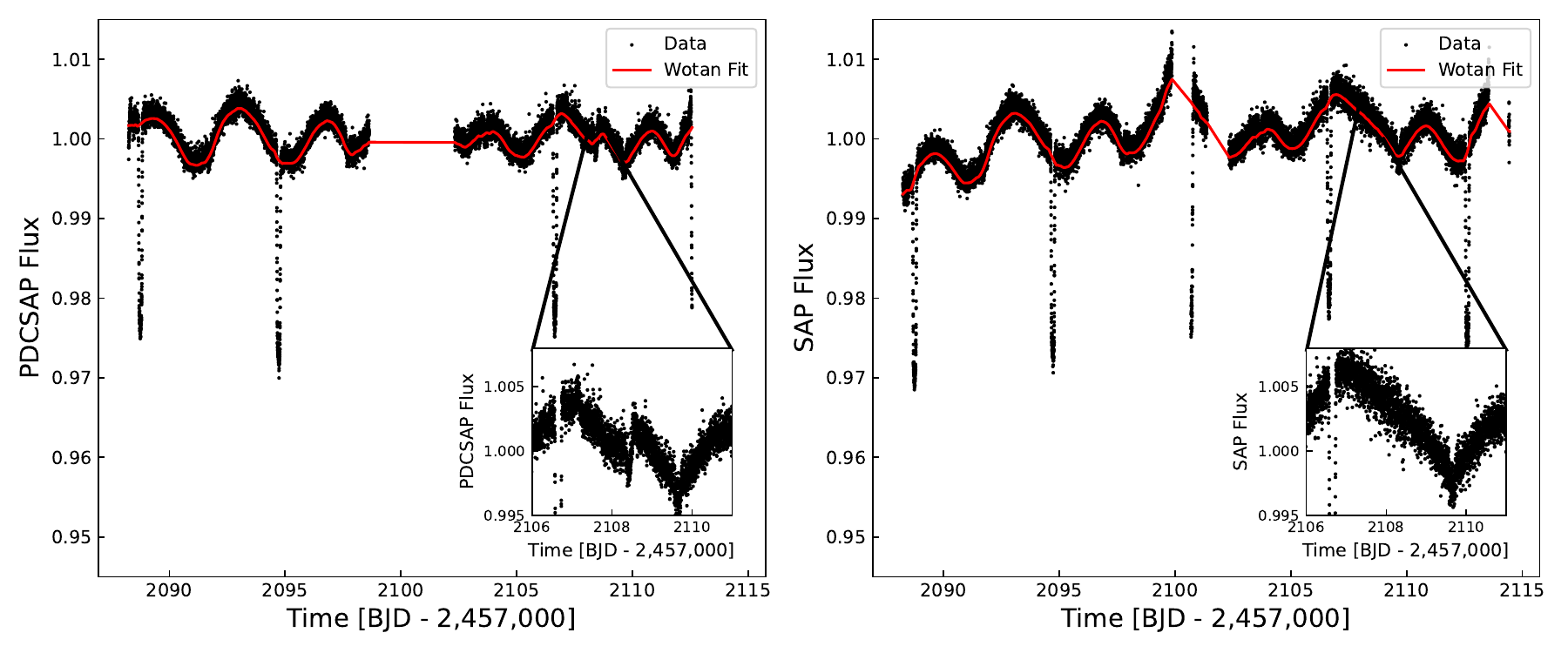} 
    \caption{PDCSAP and the SAP flux data (black points) with the \textsc{Wotan} fit (red line) that was used to de-trend the data. The zoomed-in graph in the bottom right corner of the PDCSAP plot shows an artifact that resembles another transit. The zoomed-in graph on the SAP plot shows no artifact over the same period. This suggests that the artifact came from data processing and is not an astrophysical effect. This figure also shows the sinusoidal pattern in the light curve as a result of star spots, which allowed \citet{Sethi2024} to measure the rotation period of J0021$-$16}\label{fig:PDCSAP vs SAP Transit}
\end{figure*}

EBLM J0021$-$16 is an eclipsing single-lined spectroscopic binary (SB1). The observational properties of the system are summarized in Table~\ref{Observation_table}. It was first discovered by the Wide Angle Search for Planets (WASP) and labeled as a WASP transiting hot Jupiter candidate, with a transit depth of a couple of per cent and an orbital period of $5.97$ days. Follow-up radial velocities (RVs) began in June 2009 with the \coralie ~spectrograph on the 1.2-metre Swiss Euler telescope in La Silla, Chile. \coralie ~is a fiber-fed \'{e}chelle spectrograph installed on the 1.2-m Leonard Euler telescope at the ESO La Silla Observatory, which has a resolving power of R = 50,000\,--\,60,000 \citep{Queloz2000a}. \coralie ~has a wavelength range of $390-680$ nm.

The \coralie ~RVs revealed a semi-amplitude of over 20 km/s, which is consistent with an M-dwarf in an eclipsing binary, rather than a transiting hot Jupiter. The target was therefore added to the EBLM program and was continued to be observed with \coralie ~until 2014. The RV measurements (Fig.~\ref{fig:RVs}), first published in \citet{Triaud2017}, were demonstrated to be consistent with a single Keplerian model plus a linear drift. This linear drift is likely indicative of a tertiary body. The Gaia re-normalized unit weight error (RUWE) is 1.15, whereas a value greater than 1.5 would be a stronger indication of an additional star on a wide orbit. Gaia DR4 (due mid-2026), with its point-by-point astrometry and radial velocities, may further characterize the potential third body. In Sect.~\ref{subsec:photometry} we account for the contamination by the light from an additional body. We also briefly discuss the tertiary companion in Sect.~\ref{subsec:tides} in the context of the binary's evolution. 

\citet{Triaud2017} also discovered a negative correlation between the residuals to the best-fitting radial velocity model and the spectrum bisector (Fig.~\ref{fig: Bisector Plot}). This is an indication of stellar activity, which is used to avoid false-positive exoplanet discoveries \citep{Queloz2001}. Given the strength of the spot modulation seen in TESS photometry (Fig.~\ref{fig:PDCSAP vs SAP Transit}), the bisector effect is not surprising. This bisector effect is also indicative of a tertiary body orbiting the system, causing a linear drift \citep{Triaud2017}. However, we are not able to say much on the characterization of the object with the data we have.

On September 10, 2010, spectroscopic data of EBLM J0021-16 was observed with \coralie ~during the primary eclipse to measure the RM effect. There were 33 observations taken between 02:08 UT and 09:32 UT, each with a 600-second exposure time. These were in addition to the 34 RVs taken out of eclipse to model the Keplerian orbit. All data were processed using the standard \coralie ~data reduction pipeline \citep{Baranne1996} using the cross-correlation method.

\subsection{TESS Photometry}\label{subsec:photometry}

EBLM J0021$-$16's photometry was recorded in TESS sector 29 (August to September 2020) and is scheduled for another observation in August 2026. We downloaded both the SAP (Simple Aperture Photometry) and PDCSAP (Pre-search Data Conditioned Simple Aperture Photometry) light curves via \textsc{lightkurve} \citep{lightkurve2018}. A comparison between the two is shown in Fig. \ref{fig:PDCSAP vs SAP Transit}. It is often advantageous to use PDCSAP flux because it accounts for some instrumental effects, has some cosmic ray mitigation, and is corrected for dilution from nearby stars. However, we chose to use SAP flux for two reasons. First, we noticed what looked like a transit signature from a third body (a would-be circumbinary planet, shortly after day 2108 BJD - 2,757,000). However, no such signal was seen in the SAP data. This indicates that the ``transit" was likely a processing error. Second, the SAP flux has significantly more data compared to the PDCSAP flux. The SAP flux has a partial primary eclipse at $\sim$2100 (BJD - 2,757,000) and another full primary eclipse at $\sim$2113 (BJD - 2,757,000). The SAP flux does have a slight curvature to it across the TESS sector (which comprises of two 14-day orbits of the TESS spacecraft around Earth), but other than that, it does not seem to suffer from significant instrumental effects. Both the SAP and PDCSAP data have the starspot rotational modulation.

The TESS Input Catalog (TIC) reports a contamination ratio of approximately 0.4\% to the light curve of EBLM J0021$-$16 calculated via the method outlined in \citet{StassunTESS}. We note that of all neighbors to J0021$-$16, only a single star, TIC 406352712, has the proximity and magnitude to significantly contaminate our data, being 26.4" away with a magnitude difference of $\Delta G = 5.5$. Our RVs also indicated a linear trend, likely indicative of a bound tertiary companion that may also contaminate our data. Contamination is accounted for by default in the PDCSAP flux, but not in the SAP flux. To ensure that we properly account for contamination, we compared the depth of eclipses for the system obtained from the PDCSAP and SAP data. To take into account the contamination, we divide by the CROWDSAP, or \textit{crowding factor}, which takes into account the crowding around a star in a given pixel. Once the crowding is accounted for in the SAP flux, we compared the Box Least Squares (BLS) estimated depths and found that the depths yielded 2.4609\% for the PDCSAP light curve and 2.4506\%, which is equivalent to a 0.4\% dilution rate. This is consistent with the reported TESS contamination ratio. Hence, we conclude that the contamination to the TESS photometry of J0021$-$16 has been appropriately accounted for.

\section{Methods}\label{section:methods}

\subsection{Primary Star Parameters Using \textsc{Exofastv2}}

\begin{figure}
    \includegraphics[width=0.49\textwidth]{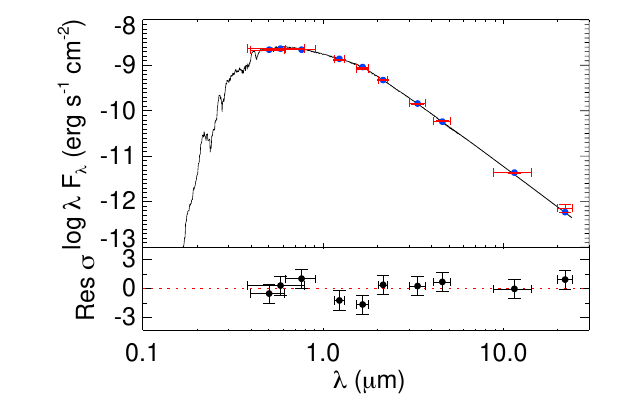}
    \includegraphics[width=0.45\textwidth]{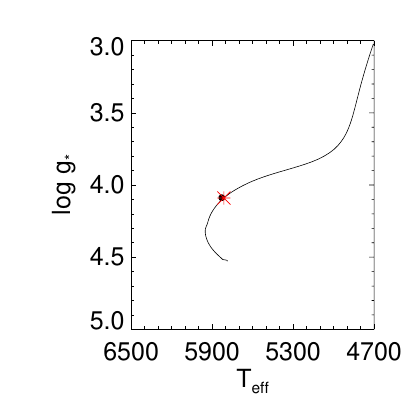}
    \caption{SED (top) and evolutionary track (bottom) of EBLM J0021-16 found by \textsc{Exofastv2}. In the bottom graph, the black line represents the evolutionary track for the best fit stellar mass. The black dot represents the $\log g$ and $T_{\rm eff}$ from the \textsc{Exofastv2} fit. The red asterisk represents the predicted value by the evolutionary track. The MIST model indicates an age of $8.5 \pm 2.8$ Gyr.}
    \label{fig: SED_Graph} 
\end{figure}

 We use \textsc{Exofastv2} \citep{Eastman2019} to fit a spectral energy distribution (SED) to constrain the radius and effective temperature of the primary star. The observed broadband photometric data are taken from Gaia, 2MASS and WISE. Fig. \ref{fig: SED_Graph} shows the SED model generated for EBLM J0021-16A. Note that the bolometric flux ratio between the primary and secondary stars is $F_{\rm B}/F_{\rm A} = (R_{\rm B}/R_{\rm A})^2\times (T_{\rm eff, B}/T_{\rm eff, A})^4 \approx 0.0015$, and hence a single-star SED fit is reasonable.

As EBLM J0021-16 is a single-lined spectroscopic binary, we cannot directly measure both stellar masses in a model-independent way. We use \textsc{Exofastv2} to fit a MESA Isochrone and Stellar Tracks (MIST) model \citep{Dotter2016} to estimate the mass of EBLM J0021-16A, as was done in previous EBLM work \citep{Duck2023}. The stellar evolutionary track for our system is shown in the bottom panel of Fig. \ref{fig: SED_Graph}. From our analysis, we find that EBLM J0021-16A is similar to our Sun but slightly more evolved, with $M_{\rm A}=1.05M_\odot$, $R_{\rm A}=1.44R_{\odot}$ and an age of $8.5\pm2.8$ Gyr. This value of $M_{\rm A}$ is ultimately used as the prior in our joint fit to further characterize the primary star's mass, as the combination of RVs and eclipses provides a strong constraint on the primary star's density.

 \subsection{Light Curve Detrending} 
\begin{figure*}
    \includegraphics[width=0.95\textwidth]{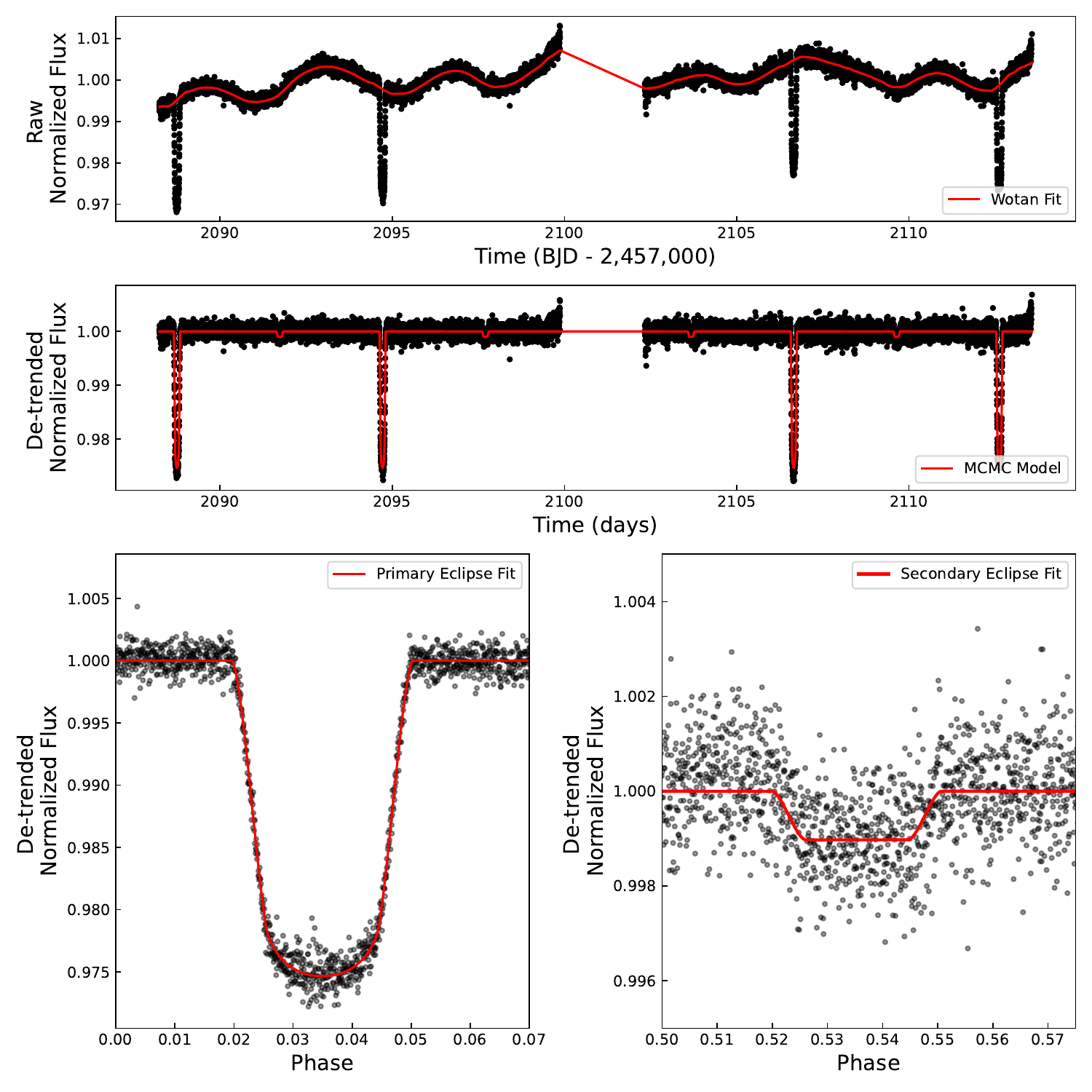}
    \caption{Raw (top) and detrended (middle) light curves of EBLM J0021-16. The bottom left plot shows the phase-folded primary eclipse and the MCMC fit, and the bottom right graph shows the phase-folded secondary eclipse and the MCMC fit.}\label{fig: Eclipses and Light curves}
\end{figure*}

We detrend the TESS light curve using the \textsc{Wotan} \citep{Hippke2019}
 Python package to remove any non-eclipse variability. This includes the effects of spot modulation on the primary star, gyroscopic movements from TESS, and scattered light by the Earth. We apply a Tukey's biweight filter with the window length of 0.7 days to ensure that the eclipse depths and shapes are preserved while other activity is removed. The raw and detrended light curves are shown in Fig. \ref{fig: Eclipses and Light curves}.




\subsection{Spectroscopy and Photometry Model Fit}


\subsubsection{\textsc{Pymc3} and \textsc{Exoplanet}}\label{subsubsec:Fitting RVs}

To determine the physical properties and orbital parameters of both stars in the binary, we use the \textsc{Exoplanet} \citep{exoplanet:joss} package to model the eclipse light curves and RV profile, and employ the MCMC algorithm to derive the posterior of each property. \textsc{Exoplanet} contains a suite of functionality that includes solving Kepler's equation, \citet{Mandel2002} light curves, quadratic limb darkening parameterization, and specialized impact parameter priors. Although originally developed for modeling stars with exoplanets, \textsc{Exoplanet} is perfectly applicable to eclipsing binaries and has been used in past studies including \citet{Pass2023} and \citet{Martin2024Outliers,Martin2024CMDra}.

The \textsc{Pymc3} package \citep{pymc3} is then used in conjunction with \textsc{Exoplanet} to determine the best-fit parameters via the MCMC method. We use the built-in optimizer to find the \textit{maximum a posteriori (MAP)} set of parameters and use it as the starting point of the MCMC sampling. The optimizer utilizes the Limited-memory Broyden–Fletcher–Goldfarb–Shanno (L-BFGS-B; \citealt{LMBFGS}) algorithm. The default fitting algorithm is the No-U-Turn Sampler (NUTS; \citealt{NUTS}), which we use for all of our fits.

\subsubsection{Fitting Keplerian RVs} \label{RV Fitting}

We first perform a fit solely based on the Keplerian RVs model, i.e. in-eclipse RVs were ignored. We fit the RV orbit with the following parameters: RV semi-amplitude ($K$), orbital period ($P_{\rm orb}$), time of eclipse ($T_{\rm 0}$), eccentricity ($e$) and the argument of periastron ($\omega$). Additionally, we fit the systemic radial velocity ($\gamma$) and a background radial velocity linear trend ($\dot{\gamma}$), which was shown by \cite{Triaud2017} to be a necessary addition to the fit, and is likely indicative of a long-period third body in the system--most likely a third star. We adopt the priors from \cite{Triaud2017} for the period and semi-amplitude. Priors for the time of eclipse and systematic background radial velocity were found by taking the median of the observation times and the mean of the RV data, respectively. The time of eclipse, systematic background radial velocity, $\log(K)$, and $\log(P_{\rm orb})$ were sampled using a normal distribution with a mean around their priors and a reasonable standard deviation. We sample over $\log(K)$ and $\log(P_{\rm orb})$ instead of the semi-amplitude and period to more efficiently sample the parameter space. For $e$ and $\omega$, rather than sampling for these two parameters directly, we opt to sample $h=\sqrt{e}\sin\omega$ and $k=\sqrt{e}\cos\omega$ to avoid introducing a biased estimate for $e$. 



We use \textsc{exoplanet} to model the radial velocity of a Keplerian orbit. We then optimize the model parameters to find the MAP model and perform an MCMC with \textsc{Pymc3}. The resulting RV parameters from this model are then used as priors for the joint RV and photometry fit. It is important to note that we measure an RV linear trend of $\dot{\gamma}=-28.7\pm0.80 \rm{m/s/yr}$, which consistent with values from \citet{Triaud2017}. Because the linear trend is so small, and no higher order trend is needed to fit the system, we are not able to make any inferences on the tertiary body that may be causing this trend.

\subsubsection{Joint RV and Photometry Fit}

After getting the best-fit parameters from the Keplerian RV fit, we create a joint fit for the RV and photometry data sets. The posterior values from the RV fit are taken as priors for the parameters mentioned in Section \ref{RV Fitting}. Additional parameters in the joint fit include the mass and radius of the primary star ($M_{\rm A}$, $R_{\rm A}$), the mass and radius of the secondary star ($M_{\rm B}$, $R_{\rm B}$), impact parameter ($b$), and quadratic limb darkening coefficients of both stars. To fit for the mass and radius of the primary star, we take the values from the \textsc{Exofastv2} SED fit as our priors. The secondary star's mass and radius are calculated using the primary star's parameters, multiplied by the mass ratio ($q=M_{\rm B}/M_{\rm A}$) and radius ratio ($R_{\rm B}/R_{\rm A}$) respectively. The mass ratio is sampled on a logarithmic scale, while the radius ratio is sampled on a linear scale. We sample the impact parameter ($b$) using the built-in \textsc{exoplanet} function \textit{xo.distributions.ImpactParameter}, which samples from a uniform distribution between zero to $1+R_{\rm A}/R_{\rm B}$ \citep{exoplanet:joss}; this range is given in the prior distribution in Table \ref{table:MCMC Params}. 

The primary and secondary eclipse light curves are modeled using \textit{xo.SecondaryEclipseLightCurve} from \textsc{Starry} \citep{Luger2019}. The quadratic limb darkening coefficients are fitted using the \textit{xo.QuadLimbDark} function. 

Similar to the RV fit, we optimize the starting parameters, get  MAP values, and then perform the MCMC sampling. We then use the posteriors from our joint fit as the priors to fit for the RM anomaly.


\begin{figure}
    \includegraphics[width=0.49\textwidth,trim={15cm 5cm 15cm 4cm},clip]{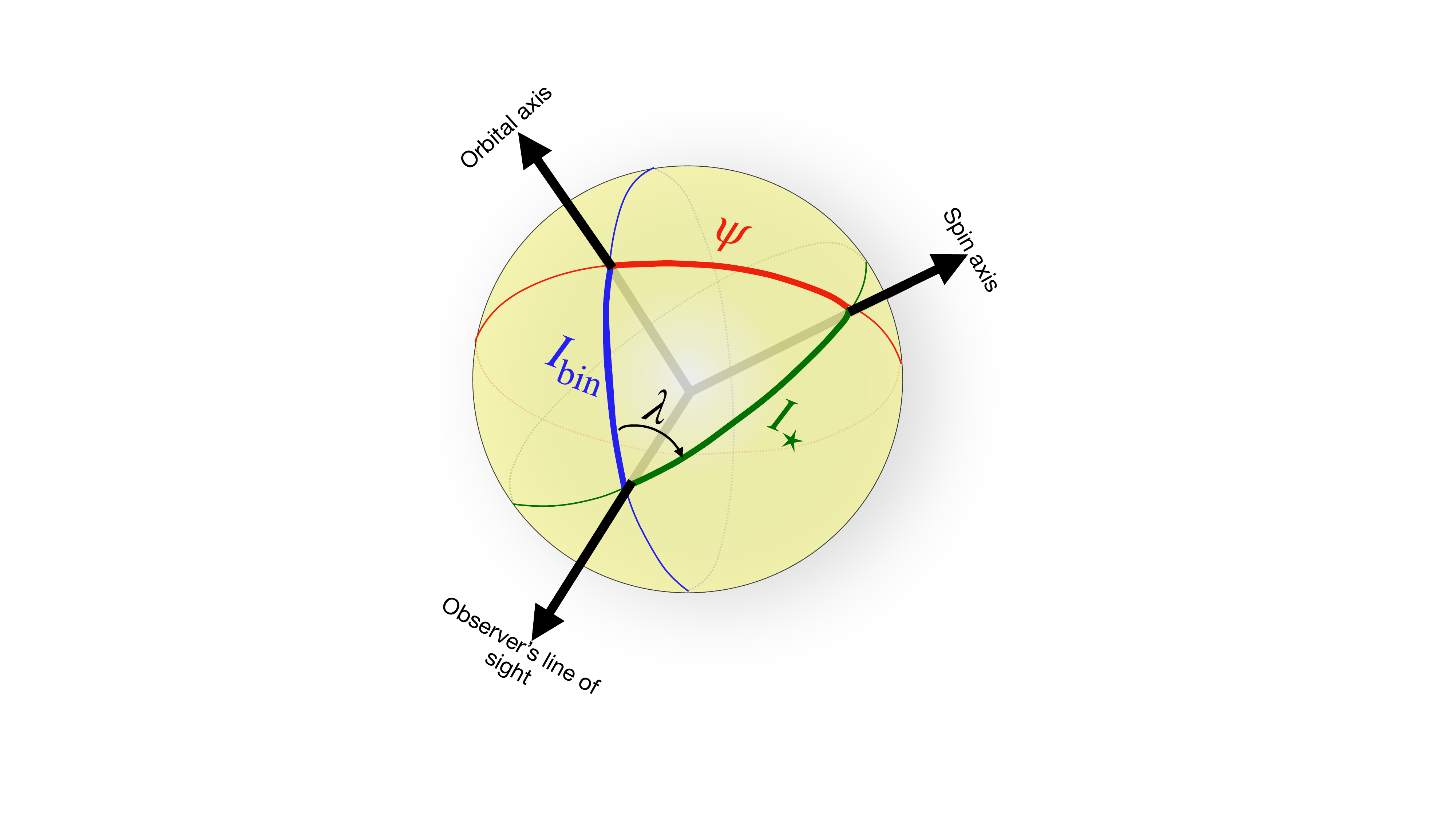}
    \caption{Orbital geometry of stellar obliquities adapted from a figure in \citet{Albrecht2021}. The diagram shows $\psi$ - the true stellar obliquity, $\lambda$ - the sky-projected spin-orbit angle, $I_{\rm \star}$ - the inclination of the rotation axis of the primary star, and $I_{\rm bin}$ - the orbital inclination of the binary}.\label{fig: Orbital Geo}
\end{figure}

\begin{figure}
    \includegraphics[width=0.49\textwidth,trim={3cm 1cm 4cm 1cm},clip]{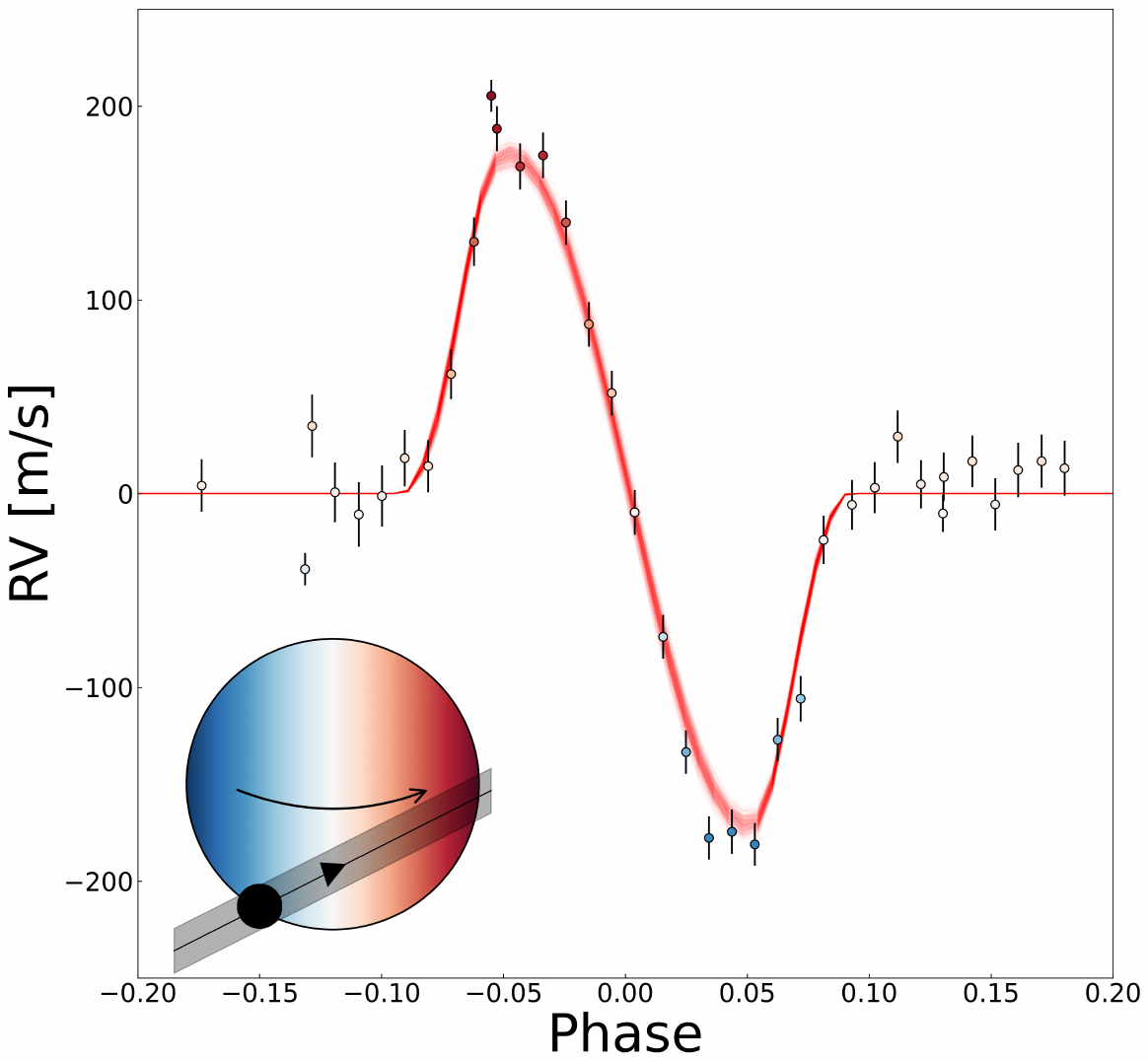}
    \caption{Fit of the RM effect for J0016$-$21. The points are colored based on the Doppler shift occurring during the primary eclipse. The diagram in the bottom left corner gives a visual representation of the orbit with an impact parameter of $b = 0.6021\pm0.0087$ and a true obliquity of $\psi = 28.9\pm2.1^{\circ}$. While the fit of the RM effect only gives us the sky-projected obliquity of $|\lambda| = 2.0\pm1.1^{\circ}$, we could find the true obliquity using the rotation period of the system. As the secondary star is blocking the blue light from the primary star, the radial velocity points get red-shifted, and as the red light is being blocked, the radial velocity points get blue-shifted.} \label{fig: RM Example Graph}
\end{figure}

\subsubsection{Secondary Star Effective Temperature}

The depth of the secondary eclipse (Fig.~\ref{fig: Eclipses and Light curves}, bottom right) is indicative of the surface brightness ratio ($S_{\rm B}/S_{\rm A}$) between the two stars. This is not quite the same as the effective temperature ratio. Since we are only observing the eclipses in one photometric bandpass, and hence have no colour information, we have to be creative to estimate $T_{\rm eff, B}$. We follow the same procedure as done previously by \citet{Swayne2020,Swayne2021,Canas2022, Duck2023}. The \textit{xo.SecondaryEclipseLightCurve} function in our \textsc{exoplanet}/\textsc{starry} MCMC directly constrains the stellar radius ratio ($k=R_{\rm B}/R_{\rm A}$) and the secondary eclipse depth ($D_{\rm sec}$). These are related to $T_{\rm eff, B}$ via

\begin{align}
    \label{eq:sec_eclipse_depth}
    D_{\rm sec} &=k^2S + A_{\rm g} \left(\frac{R_{\rm B}}{a}\right)^2, \\
    \label{eq:sec_eclipse_depth2}
    &= \left(\frac{R_{\rm B}}{R_{\rm A}}\right)^2\frac{\int \tau(\lambda)F_{\rm B,\nu}(\lambda,T_{\rm eff,B})\lambda d\lambda}{\int \tau(\lambda)F_{\rm A,\nu}(\lambda,T_{\rm eff,A})\lambda d\lambda} + A_{\rm g} \left(\frac{R_{\rm B}}{a}\right)^2,
\end{align}
where $A_{\rm g}$ is the geometric albedo, $\tau(\lambda)$\footnote{\url{http://svo2.cab.inta-csic.es/svo/theory/fps3/index.php?id=Kepler/Kepler.K}} is the TESS transmission function as a function of wavelength $\lambda$ and $F$ is the normalised flux. For eclipsing M-dwarfs, $A_{\rm g}$ is close enough to zero and the secondary star temperature is hot enough such that the second term of Eq.~\ref{eq:sec_eclipse_depth2} can be ignored.

We calculate the integrated flux of the primary star using a high-resolution PHOENIX model spectrum, convolved with the TESS bandpass. The model spectrum is based on the  best-fit values for $T_{\rm eff,A}$, $\log g_{\rm A}$ and [Fe/H] from \textsc{EXOFASTv2}. We then calculate the brightness of the secondary star across a grid of potential effective temperatures between 2300 and 4000 K. Using Brent minimisation we solve for $T_{\rm eff,B}$. This process is repeated with values of the radius ratio ($R_{\rm B}/R_{\rm A}$) and secondary eclipse depth ($D_{\rm sec}$) randomly drawn from $1\sigma$ errors of the MCMC-fitted parameters. From this we derive a posterior and errorbar on $T_{\rm eff,B}$.

\begin{figure*}
    \includegraphics[width=0.99\textwidth]{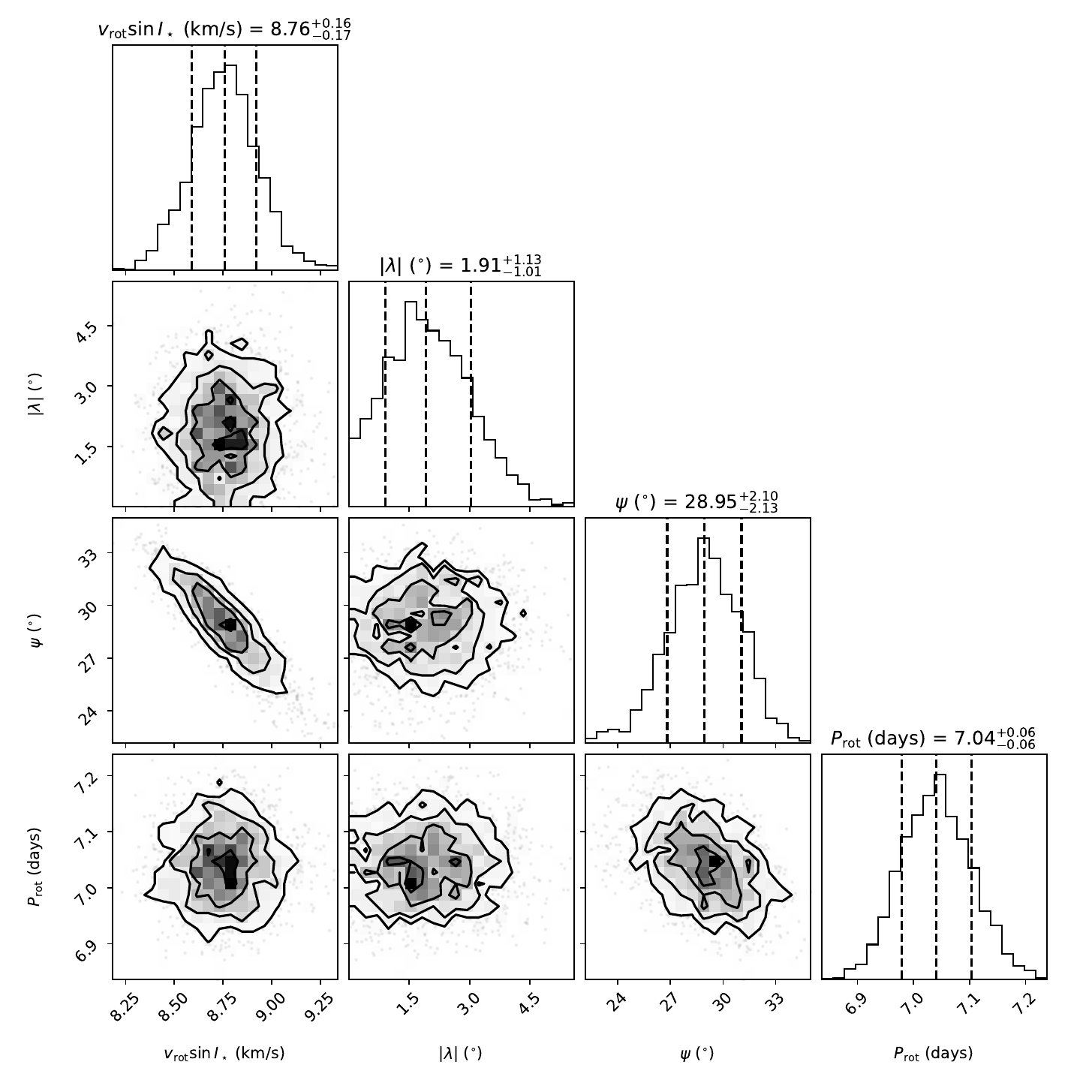}
    \caption{Corner plot produced from the MCMC simulation of the joint fit to all RVs (Keplerian and Rossiter) and the TESS photometry for primary and secondary eclipses. The plot shows the correlation between the parameters corresponding to the RM fit - $v_{\rm rot} \sin I_{\rm star}$, $\lambda$, $\psi$, \& $P_{\rm rot}$.}\label{fig: Corner Plot 4 Params}
\end{figure*}

\subsection{Fitting the Rossiter-McLaughlin Anomaly} \label{subsection:RM Fit}

We use the RM effect to measure the alignment between the stellar spin axis and the orbital axis in EBLM J0021-16. Since EBLM has a very uneven flux ratio ($F_{\rm B}/F_{\rm A}\sim0.0015$), we are inherently referring to the spin-axis of the primary star. At its core, the RM effect is a distortion of stellar spectral lines during a transit or eclipse, caused by the foreground object (in our case, EBLM J0021-16B) blocking rotating regions of the background star (EBLM J0021-16A) and creating an asymmetric Doppler profile. Traditional methods for RM fitting constrain the projected rotational velocity of the occulted star, $v_{\rm rot} \sin I_{\rm \star}$, and the magnitude of the projected spin-orbit obliquity, $|\lambda|$. \citet{Hirano2011} updated RM models to account for macroturbulence ($\zeta$), thermal broadening ($\beta$) and Lorentzian broadening (natural and pressure ($\gamma$). We combine the \citet{Gupta:2024} publicly-available implementation of the \citet{Hirano2011} model into our \textsc{exoplanet} joint fit of the Keplerian RVs and photometric eclipses. This is similar to what \citet{Carmichael2025} used to fit the RM effect of brown dwarfs with {\tt \textsc{EXOFASTv2}}. In Fig. \ref{fig: RM Example Graph} we show the fitted RM anomaly.

The Rossiter-McLaughin effect constrains the \textit{sky-projected obliquity, $\lambda$}, which is the angle between the sky projections of the orbital and rotational axes. This tells us how misaligned our system is as seen from Earth. This is different from the true obliquity, $\psi$, which is measured between the axes in three dimensions \citep{Albrecht2021}. The relationship between the two obliquities is demonstrated in Fig.~\ref{fig: Orbital Geo} and given by 

\begin{equation}
    \cos \psi = \cos I_\star \cos I_{\rm bin} + \sin I_\star \sin I_{\rm bin} \cos \lambda,
\end{equation}
where $I_{\rm bin}$ is the orbital inclination of the binary and $I_\star$ is the inclination of the rotation axis of the primary star. 

For most systems, the projected obliquity ($\lambda$) is the only obliquity that can be measured. However, if a star has a known rotational period, we can calculate the true obliquity ($\psi$). This comes from constraining $\sin I_\star$ via two independent methods. First, the RM fit provides a constraint on the projected rotation rate of the primary star, $v_{\rm rot}\sin I_\star$. Second, if the true rotation period of the primary star, $P_{\rm rot}$, is measured from photometry, then we get the true rotation speed as $v_{\rm rot} =2\pi R_{\rm A}/P_{\rm rot}$. Trivially then,

\begin{equation}
    \sin I_{\rm \star} = \frac{v_{\rm rot} \sin I_{\rm \star}}{v_{\rm rot}}.
\end{equation}
By constraining $\sin I_\star$, and also using the fact that it's an eclipsing binary, so $I_{\rm bin}\approx 90^{\circ}$, we obtain an equation for the true obliquity,

\begin{equation} \label{True Obl}
    \cos\psi \approx \cos\lambda \frac{P_{\rm rot}v_{\rm rot}\sin I_{\rm \star}}{2\pi R_{\rm A}}.
\end{equation}

We found EBLM J0021-16 to have a stellar inclination of $I_{\rm \star}=57.9\pm2.1^{\circ}$. The stellar rotation period of EBLM J0021-16A was measured by \citet{Sethi2024} using spot modulations. Since star spots are cooler and dimmer than the surrounding photosphere, the presence of a spot overdensity leads to periodic variations in brightness as the star rotates. In the case of EBLM J0021$-$16, the spot modulation is clearly visible in Fig. \ref{fig:PDCSAP vs SAP Transit}. By measuring the period of the out-of-eclipse variability in the TESS light curves, \citet{Sethi2024} determined the stellar rotation period to be $P_{\rm rot} = 7.042 \pm0.061$ days. 

We include this value of $P_{\rm rot}$ in our model as a Gaussian and use this to solve for $\sin I_\star$. We combine this with the best-fitting $\lambda$ from the RM effect and solve for $\psi$ using Eq. \ref{True Obl}. The final values for $\psi$ and $\lambda$ along with all the other parameters from the RM, RV, and photometry joint fit can be found in Table \ref{table:MCMC Params}. The distributions of $\lambda$, $\psi$, $P_{\rm rot}$, and $v_{\rm rot} \sin I_{\rm \star}$ can be seen in the corner plot in Fig. \ref{fig: Corner Plot 4 Params}.


\section{Results}\label{section:results}

Our best-fitted parameters are shown in Table \ref{table:MCMC Params}. Using the methods outlined in Section \ref{section:methods} for modeling the RM anomaly, we found a sky-projected obliquity of $|\lambda| = 2.0 \pm 1.1^{\circ}$ and a true obliquity of $\psi = 28.9\pm2.1^{\circ}$. This obliquity suggests the system is not yet spin-orbit aligned with a prograde, almost circular orbit (Fig. \ref{fig: RM Example Graph}, $e = 0.000534 \pm 0.000095$). 

There is a possible degeneracy between $v_{\rm rot} \sin I_{\rm \star}$ and $\lambda$ when the impact parameter, $b$, is close to zero \citep{Albrecht2011}. The RM effect in a system with a high $v_{\rm rot} \sin I_{\rm \star}$ and large $\lambda$ may appear similar to that of a system with a low $v_{\rm rot} \sin I_{\rm \star}$ and small $\lambda$ because the object spends the same amount of time in each hemisphere \citep{Triaud2018}. This degeneracy can be partially resolved if the stellar rotation of the system is known. For our target, we found an impact parameter of $b=0.5961 \pm 0.0089$, which is large enough that the degeneracy does not apply here.

\section{Discussion} \label{section:discussion}

\subsection{Tidal Effects}\label{subsec:tides}

\begin{table}
\caption{Orbital period ($P_{\rm orb}$), secondary mass ($M_{B}$), projected obliquity $|\lambda|$, and true obliquity ($\psi$, if measured) for all EBLM targets with measured RMs. This includes the brown dwarf WASP-30.}              
\label{EBLM Target Comp}      
\centering                                    
\begin{tabular}{m{1.15cm} p{1.4cm} p{0.9cm} p{0.8cm} p{0.5cm} p{1.4cm}}

\hline\hline                        
 
 Target & $P_{\rm orb}$ & $M_{B}$ & $|\lambda|$ & $\psi$ & Citation\\

  & [days] & [$M_{\odot}$] & [$^\circ$] & [$^\circ$] & \\
 
\hline 

WASP-30 & 4.156739 $\pm0.000011$ & 0.0597 $\pm0.0011$ & 7 $\pm23$ & & \citet{Triaud2013} \\
EBLM J1219-39 & 6.7600098 $\pm0.000028$ & 0.0910 $\pm0.0021$ & 4.1 $\pm5$ & & \citet{Triaud2013} \\
EBLM J0218-31 & 8.884102 $\pm0.000011$ & 0.390 $\pm0.009$ & 4 $\pm7$ & & \citet{Gill2019} \\
EBLM J0608-59 & 14.608559 $\pm0.000013$ & 0.313 $\pm0.011$ & 2.8 $\pm17.1$ & & \citet{hodzic2020} \\
EBLM J0021-16 & 5.96727088 $\pm0.00000034$ & 0.1877 $\pm0.0011$ & 2.0 $\pm1.1$ & 28.9 $\pm2.1$ & Spejcher et al. (this work) \\

\hline

\end{tabular}

\vspace{1mm}
{\raggedright 
\small \textit{Note ---} \citet{Triaud2013} accounted for asymmetrical errors in their fits of WASP-30 and EBLM J1219-39, so the errors given here are the average of the asymmetric errors given in \citet{Triaud2013}. This was done to provide a rough symmetry of errors to match the errors from our calculation, which assumes a $1\sigma$ symmetry.\par}

\end{table}

\begin{figure*}
    \includegraphics[width=0.49\textwidth]{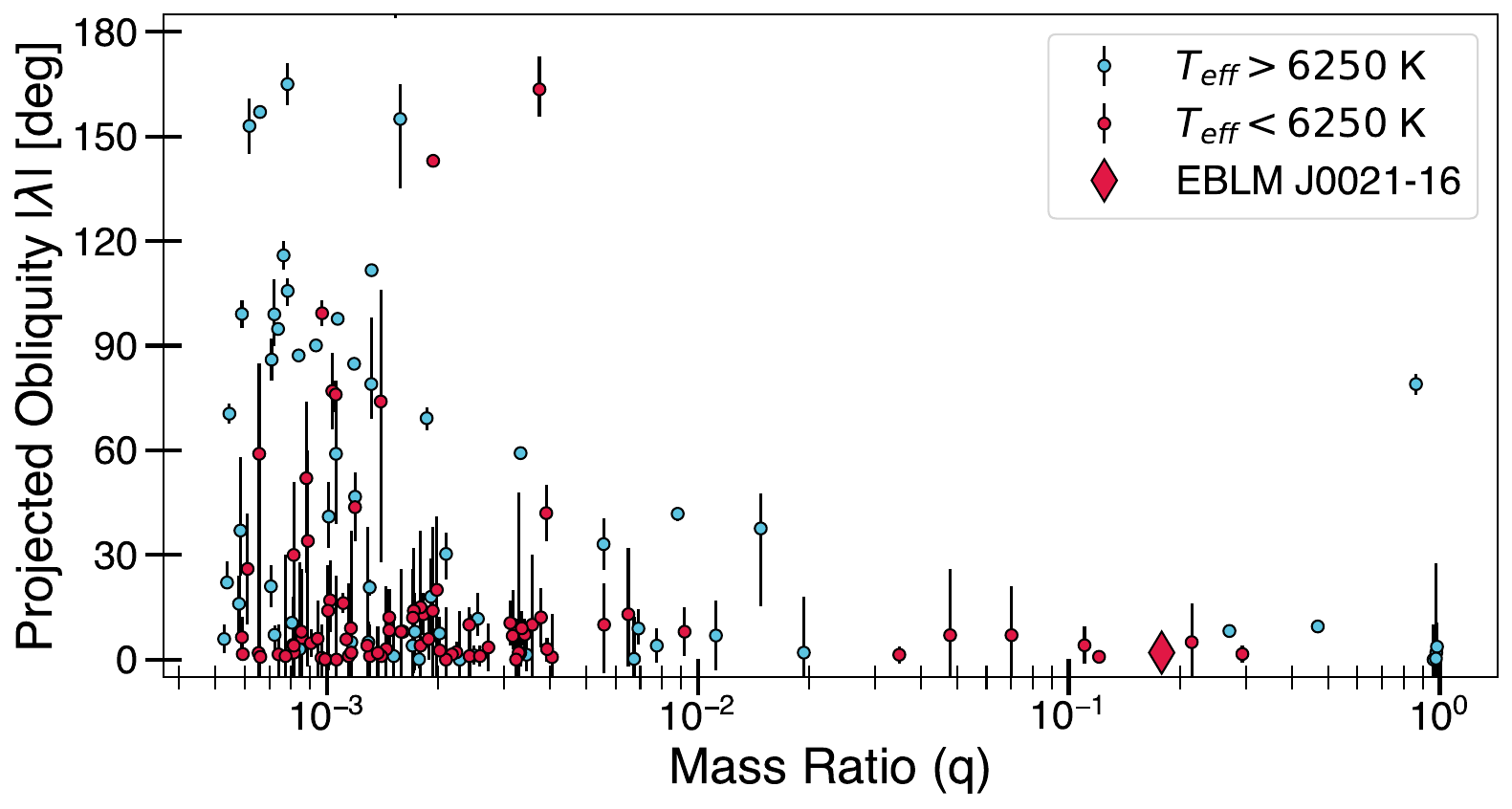}
    \includegraphics[width=0.49\textwidth]{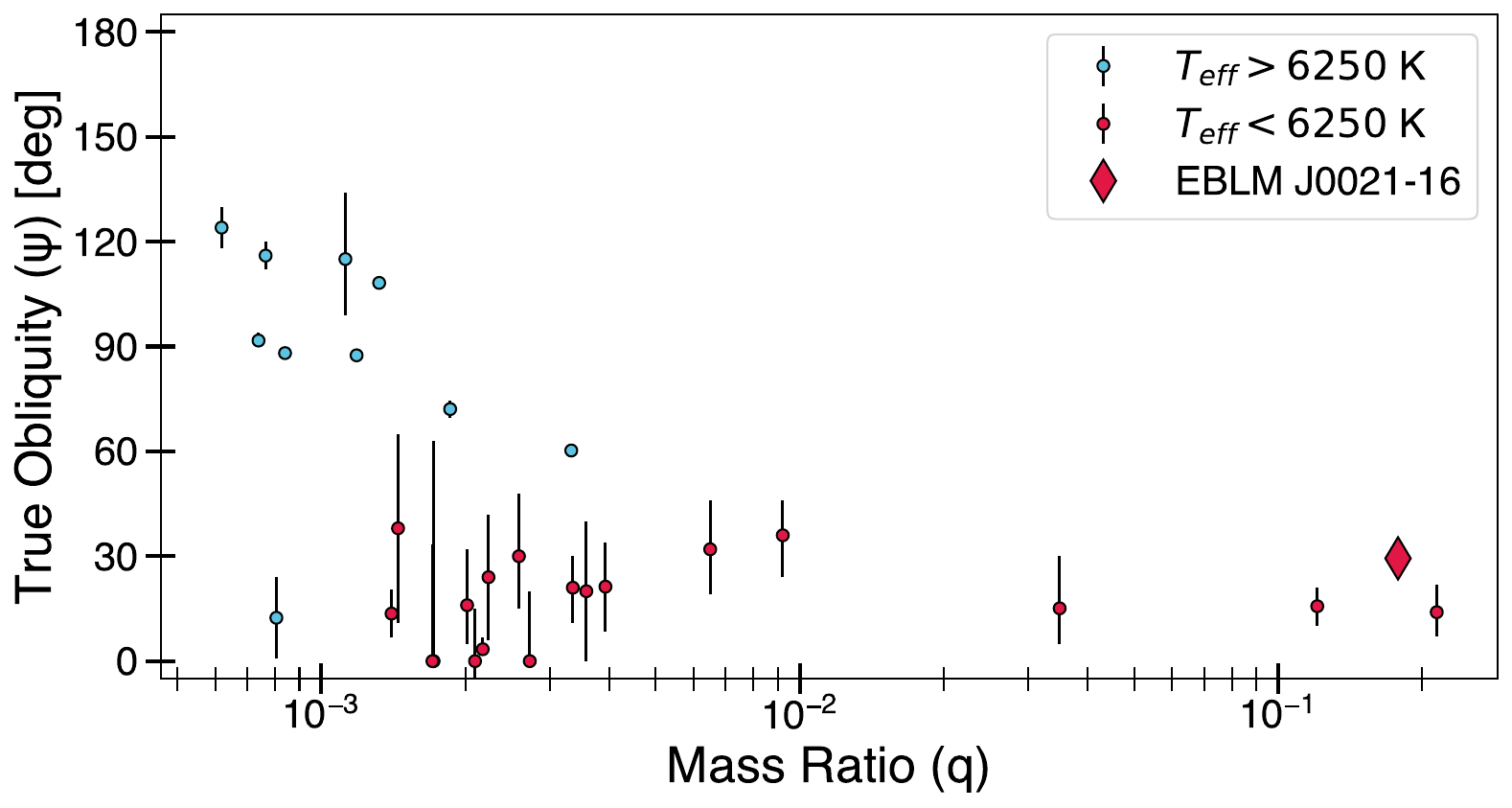}
    \caption{Sky-projected stellar obliquity $\rm |\lambda|$ vs. mass ratio (q, left) and true 3-D Stellar Obliquities ($\psi$) vs mass ratio (right) for eclipsing companions, $M_b > 0.7$ $M_{\rm J}$. Obliquities measured via apsidal motion are not depicted. Red points depict systems with primary stars below the Kraft break \citep[$\sim6250$ $K$;][]{Kraft1967}, and blue points depict those above. EBLM J0021-16 is shown as the red diamond. The giant planet and brown dwarf obliquities were retrieved from TEPCat \citep{Southworth2011} and \citet{Carmichael2025}. The eclipsing binary obliquities are from \citet{Marcussen2022} and \citet{Wells2025}.}
    \label{fig:lambda-q}
\end{figure*}

For binary stars with sufficient angular momentum, tides typically act to bring the system to three equilibrium states, so long as they have had sufficient time to do so \citep[e.g.][]{Hut1980,O2014,Barker2025}:
\begin{enumerate}
    \item spin-orbit alignment: the true obliquity $\psi=0^{\circ}$;
    \item spin-orbit synchronization: the stellar rotation period 
    equals the orbital period, $P_{\rm rot}=P_{\rm orb}$;
    \item circularization: the orbital eccentricity $e=0$.
\end{enumerate}
In this paper, we measured $\psi= 28.9\pm2.1^{\circ}$, so EBLM J0021-16 is not quite spin-orbit aligned. Recall that we can only speak about 
the spin of the primary star. Theoretically, we can calculate the realignment timescale using \citep[e.g.][]{Lai2012,O2014,B2016}
\begin{equation} \label{Realignment Eq}
    \tau_{\psi} \approx \frac{4}{9} \frac{Q'_{2, 1, 0}r_{g}^2}{\pi} \left(\frac{M_A+M_B}{M_B}\right)^2 \frac{P_{\rm orb}^4}{P_{\rm dyn}^2 P_{\rm rot}},
\end{equation}
where $P_{\rm dyn}=2\pi\sqrt{R_{\rm A}^3/(GM_{\rm A})}$ is the dynamical timescale of the primary star and $r_g^2$ is its squared radius of gyration \citep[see][for further details]{Barker2020}. 
We have assumed that the tidal component dominating the obliquity evolution is the one with a harmonic degree $l=2$ (quadrupolar) and $m=1$ (azimuthal wavenumber), which has a corresponding modified tidal quality factor $Q'_{2,1,0}$ \citep[e.g.][]{LO2017}, and that the orbital angular momentum exceeds the rotational angular momentum. 
We compute $Q'_{2,1,0}$ by assuming that the dominant tidal mechanism driving alignment is the dissipation of inertial waves in the convective envelope of the primary \citep[motivated by][]{LO2017}, and that this is well-approximated by the value representing the frequency-averaged dissipation ($Q'$) in Section 3.1 of \citet{Barker2020} (building upon \citealt{O2013}). This provides a value of $Q'$ representing the typical level of dissipation due to inertial waves, which fully accounts for the stellar structure. While this means that we are strictly calculating $Q'$ for a different component of the tide with $l=m=2$, the values obtained from this calculation appear to fairly represent the numerical calculations for obliquity tides in \citet[][under the same assumptions]{LO2017}, but this approach allows for more straightforward calculations in stellar evolutionary models. More realistic frequency-dependent models in this system may be worthwhile in the future \citep[e.g.][]{AB2023,Dewberry2024}.

A typical value obtained for a 1.05$M_\odot$ main-sequence star is (to within a factor of 2) $Q'\approx 10^7\left(P_{\mathrm{rot}}/10 \mathrm{d}\right)^2$ \citep{Barker2020}, though the numerical value obtained for the primary star using {\sc MESA} stellar models \citep[][]{Paxton2011, Paxton2013, Paxton2015, Paxton2018, Paxton2019} (with MIST parameters in \citealt{Dotter,Choi}) consistent with its current age and rotation is $Q'\approx 1.8\times 10^6$. Using stellar models of EBLM J0021–16 A leads to $\tau_\psi\approx 10-29\, \mathrm{Myr}$ (depending on age).

Based on MIST evolutionary tracks fitted by \textsc{Exofastv2}, our rough estimate of the age of EBLM J0021–16 was $8.5\pm 2.8$ Gyr. Therefore, the binary is theoretically expected to be spin-orbit aligned. 

We can similarly compute the spin-orbit synchronization timescale using the models from \citet{Barker2020}. This is given by:
\begin{equation} \label{Synchronization Eq}
    \tau_{\rm synch} = \frac{2Q'r_{g}^{2}}{9\pi}\left(\frac{M_{\rm A} + M_{\rm B}}{M_{\rm B}}\right)^{2}\frac{P_{\rm orb}^4}{P_{\rm dyn}^{2}P_{\rm rot}}=\frac{\tau_\psi}{2},
\end{equation}
where all quantities have been defined above with the same $Q'$ used, which is more directly appropriate for $l=m=2$ synchronization tides. 
The synchronization timescale was found to be $\tau_{\rm sync} \approx 3-5$ Myr for the primary star, and hence, we expect theoretically that this system should be tidally locked. However, this is not the case; the stellar rotation period is $P_{\rm rot} = 7.042 \pm 0.060$ days, compared to the orbital period of $P_{\rm orb} = 5.967$ days. This deviation may be 
related to uncertainties in tidal physics, making the system an intriguing candidate for future studies. On the other hand, the deviation could also be attributed to stellar differential rotation \citep[e.g.][]{Balona2016,Lurie2017L}. The rotation periods derived from spot modulations, as done by \citet{Sethi2024}, may reflect rotation at higher latitudes rather than the equator. If tidal synchronization occurred at the equator, a spot-dominated higher-latitude region could yield a longer rotation period and thus mimic a non-synchronized state.

Similarly, Sethi et. al. (in prep.) calculated the circularization timescale driven by primary star dissipation following \citet{Barker2020,B2022}:
\begin{equation} \label{Circularization Eq}
    \tau_{\rm circ} = \frac{2Q'}{63\pi}\left(\frac{M_{\rm A}}{M_{\rm B}}\right)\left(\frac{M_{\rm A} + M_{\rm B}}{M_{\rm A}}\right)^{\frac{5}{3}}\frac{P_{\rm orb}^{\frac{13}{3}}}{P_{\rm dyn}^{\frac{10}{3}}}.
\end{equation} 
The models they used are consistent with $\tau_{\rm circ} \approx 84-188$ Myr (depending on the particular age),
and hence EBLM J0021-16 is expected to be circularized. The orbit is indeed very close to being circular, with $e=0.000550\pm0.000090$, which is thus consistent with our predictions for tidal dissipation of inertial waves \citep[see also][]{B2022}. This tiny eccentricity might be a function of the expected third star in the system (due to the linear trend in the RVs). Alternatively, it might be an artifact of our model fit and unaccounted for systematics.

 Based on the timescales calculated above, we expect EBLM J0021-16 to be spin-orbit aligned, synchronized, and have a circular orbit. However, we only observe a nearly circular orbit of the system, but spin-orbit misalignment and asynchronization. The observed misalignment could be caused by a tertiary body, however; we would need to know more about the tertiary body to determine the strength of the effect it would have on the alignment of EBLM J0016-21 \citep{Lai2018}. The exact cause for a nearly circular orbit but spin-orbit misalignment and asynchronization is left as a puzzle for tidal physics models.

\subsection{Obliquity Across the Spectrum of Mass Ratio}

We provide the orbital period ($P_{\rm orb}$), secondary mass ($M_{B}$), projected obliquity $|\lambda|$ and true obliquity ($\psi$) for each EBLM system with a measured RM, including EBLM J0021-16 and the brown dwarf WASP-30, in Table \ref{EBLM Target Comp}. The projected obliquity in the table shows that the targets are mostly spin-orbit aligned. EBLM J0021-16 is the first of the EBLM sample where we can measure the true obliquity, $\psi$, owing to the clear star spot rotation in TESS photometry (Fig.~\ref{fig:PDCSAP vs SAP Transit}), and can see that it is not as aligned as the small $\lambda$ would lead you to believe.

In Fig. \ref{fig:lambda-q}, we show the projected and true obliquities from RM measurements for all giant planets, brown dwarfs, and stellar companions with $M_{\rm b} > 0.7 M_{\rm J}$ \citep{Southworth2011,Marcussen2022,Carmichael2025}. The small mass ratios, concentrated around $q=10^{-3}$, are RM measurements of hot Jupiter obliquities. We can see that $\sim$ 1/3 of the population of hot Jupiters is misaligned \citep{Triaud2010, Winn2015}. For the systems whose primary star is cooler than the Kraft break ($\sim6250 \rm K$), which corresponds to stars with convective outer envelopes \citep{Kraft1967}, the secondary bodies tend to be more well-aligned. Above the Kraft break, stars have thinner convective envelopes, 
which is thought to be less efficient for 
tidal re-alignment, so the secondary bodies are more often misaligned. Alternatively, it has been proposed that gravity wave-driven dynamics could lead to differential rotation or spin-orbit misalignments in surface layers independently of tidal effects \citep[e.g.][]{Rogers2012}.

Another trend that emerges from looking at the obliquity of systems with different mass ratios is that systems with larger mass ratios tends to be more aligned. This is likely caused by tides being more efficient with more massive secondary bodies. Indeed, Eq.~\ref{Realignment Eq} shows that the re-alignment time for convective envelope stars scales as $\tau_{\rm align,CE}\propto1/q^2$, and hence eclipsing binaries (high $q$) might be all aligned, whereas there is a spread for hot Jupiters. \citet{Zahn1977} also shows that the hotter stars, with radiative outer envelopes, have an even stronger relationship between the alignment timescale and mass ratio with host stars above $1.6M_\odot$: $\tau_{\rm align, RE}\propto1/(q^2(1+q)^{5/6})$. 

We do encourage caution with respect to interpreting these early measurements of eclipsing binary obliquities. It is evident from Fig.~\ref{fig:lambda-q} that the field of RM measurements and stellar obliquities is much more developed for hot Jupiters than it is for eclipsing binaries. There are only 4 measured true obliquities for a mass ratio greater than 0.01, and all of them are for primary stars cooler than the Kraft break. The companions of these systems are EBLM J0021-16B ($\psi=28.9\pm2.1^{\circ}$), CWW 89Ab ($\psi=15.1^{+15.0\circ}_{-10.9}$), TOI-2119b ($\psi=15.7 \pm 5.6^{\circ}$), and LP 261-75C ($\psi = 14.0^{+7.8\circ}_{-6.7}$). The latter three companions are all transiting brown dwarfs with measured obliquities from \citet{Carmichael2025} using techniques similar to those outlined in Section \ref{subsection:RM Fit}. CWW 89 and LP 261-75 are confirmed triple systems consisting of a brown dwarf orbiting an inner M-dwarf with an outer M-dwarf companion. TOI-2119 is not a confirmed triple system, but has a Gaia RUWE of 1.93 \citep{Belokurov2020,GaiaCollaboration2016}, indicating an unresolved massive third body in the system \citep{Belokurov2020,Krolikowski2021}. These systems all have non-zero eccentricities with mostly aligned orbits. \citet{Petrovich2015} presents a plausible framework via coplanar high eccentricity migration that would explain the high eccentricity but spin-orbit alignment for CWW 89. \citet{Carmichael2025} proposes that this could also explain the similar orbits for TOI-2119 and LP 261-75. More RM measurements, including those already observed and soon to be published in the EBLM program, will shed more light on tidal re-alignment timescales.



\subsection{Mass-Radius Relationship}

One of the primary goals of the EBLM survey is the discovery and characterization of M-dwarfs in eclipsing binaries \citep{Maxted2023}. The goal is to measure their mass and radius to a precision of better than a couple of per cent. More precisely-measured M-dwarf parameters are crucial to our understanding of stellar structure and evolution models. Furthermore, the reliability of our planetary parameters is dependent on the reliability of the stellar parameters. This requires better M-dwarf models, both theoretical and empirical. 

We measure the mass and radius of EBLM J0021-16B to 0.6 and 0.8 \% uncertainties, respectively. However, we note that these are the nominal uncertainties, while the true uncertainty could be larger due to the unaccounted systematics. One source of systematic uncertainty comes from the method we use to derive the parameters of the primary star. Since EBLM J0021-16 is a single-lined spectroscopic binary, we do not get model-independent stellar masses. \citet{Duck2023} showed in two EBLM systems that systematic model uncertainty can be similar to the nominal uncertainty.

\begin{figure}
    \includegraphics[width=0.49\textwidth]{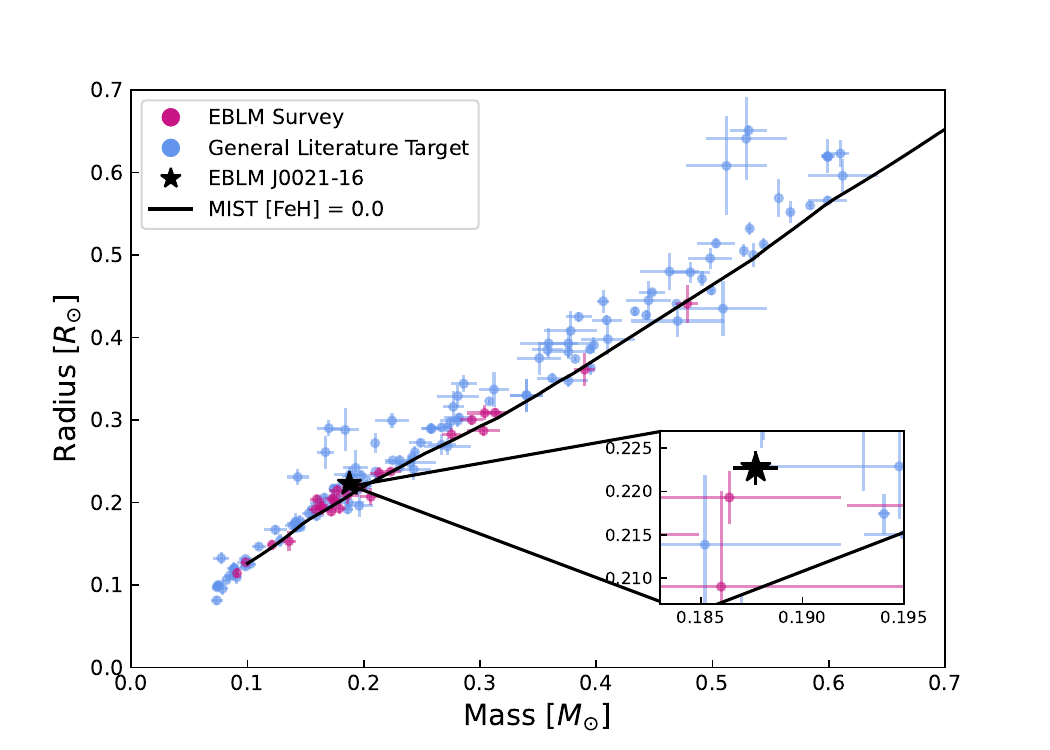}
    \caption{Mass vs. Radius plot for high-precision M-dwarfs, with data taken from \citet{Maxted2023}. The plot includes targets from the EBLM survey (pink dots) and general literature targets (blue dots). Only targets with a mass or radius better than 10\% are included here. The solid line passing through the data points indicates the MIST stellar model for [Fe/H] = 0.0. EBLM J0021$-$16B is highlighted by the black star.} \label{fig: Mass vs Radius}
\end{figure}

The distribution of mass and radius of M-dwarf population that is well-characterized, along with the mass-radius relationship obtained from MIST stellar model with [Fe/H] = 0.0, is shown in Fig. \ref{fig: Mass vs Radius}. Among this population, many are observed to have larger radii than what was expected from the model, and hence dubbed the ``radius inflation'' problem \citep{Parsons2018,Morrell2019}. EBLM J0021-16B, marked in the figure by a black star, also lies above the theoretical prediction. We note that the \textsc{Exofastv2} joint SED \& evolutionary track fit of the primary star returns [Fe/H]$=0^{+0.24}_{-0.36}$, which is consistent with Solar metallicity. To quantify the radius inflation, we use the following equation from \citet{Gill2019}:

\begin{equation} \label{Radius Inflation}
    \frac{\Delta R_{\rm B}}{R_{\rm B}} = \frac{R_{\rm B} - R_{\rm B, exp}}{R_{\rm B}},
\end{equation}
where $R_{\rm B}$ is the radius of the M-dwarf given by the MCMC fit ($R_{\rm B}=0.2214\pm0.0019R_\odot$), and $R_{\rm B, exp}$ is the expected radius value from the MIST model. To find the expected radius, we interpolate over the MIST model using the mass of the M-dwarf given by the MCMC ($M_{\rm B}=0.1877\pm0.0011$). We found the expected radius to be $R_{\rm 2, exp} = 0.2086R_{\odot}$, yielding a percentage of inflation of $\Delta R_{\rm B}/R_{\rm B} = 5.78\%$. In comparison to our derived uncertainty, EBLM J0021-16B is $6.7\sigma$ above the theoretical expectation.


\section{\MakeUppercase{CONCLUSION}}
We combined TESS photometry with 67 CORALIE RVs to characterize EBLM J0021-16, a 5.97 day eclipsing binary composed of a G-dwarf and an M-dwarf. Our RVs, taken over a span of more than 5 years, contain one series of measurements taken during a primary eclipse. Our joint photometry-spectroscopy fit provides a precisely-characterized Rossiter-McLaughlin effect, with a sky-projected spin-orbit obliquity of $|\lambda| = 2.0 \pm1.1^{\circ}$. TESS photometry reveals star spot modulations with a period of $7.042$ days \citep{Sethi2024}. This allows us to derive a true obliquity of $\psi = 28.9\pm2.1^{\circ}$. EBLM J0021-16 is precisely measured to be moderately spin-orbit misaligned on an orbit that is prograde and almost circular ($e=0.000534\pm 0.000095$). Based on the tidal theory of inertial wave dissipation, we expect the orbit to be circular and both spin-orbit aligned and synchronized. Whilst the orbit is almost perfectly circular, the lack of observed spin-orbit alignment and synchronization is a curious test for tidal physics models.

EBLM J0021-16 is the 5th RM effect published in the EBLM series, and the first to have a true 3D obliquity measurement. It is the 8th main sequence eclipsing binary to have an RM observation, and the majority have aligned (or close-to-aligned) orbits. This is in contrast to the 132 hot and warm Jupiter's that have obliquity measurements and exhibit a diversity of obliquities. This may be an expected consequence of tidal re-alignment being faster in stellar binaries than in hot Jupiters. Alternatively, tight stellar binaries might form and evolve in aligned orbits. We caution that any trends seen in the obliquitiy of eclipsing binaries are preliminary. We encourage further RM observations, particularly for eclipsing binaries with longer periods and hotter primary stars, where tides are expected to be weaker.

Finally, we measured the mass and radius of the M-dwarf, EBLM J0021-16B, to be $M_{\rm B} = 0.1877 \pm 0.0011 M_{\odot}$ and $R_{\rm B} = 0.2214 \pm 0.0019 R_{\odot}$, both to a precision better than 1\%. By comparing these values to a theoretical MIST model, we found the M-dwarf radius to be inflated by 5.78\%, adding another piece to the radius inflation puzzle.

\section*{Acknowledgements}

The EBLM Program has been supported by NASA TESS Guest Investigator Program through Cycle 1 (G011278), Cycle 2 (G022253), Cycle 3 (G03195), Cycle 4(G04157), Cycle 5 (G05073), Cycle 6 (G06022), Cycle 7 (G07005) and Cycle 8. Postage stamp observations were provided in all 7 cycles and funding was provided in cycles 2, 4, 5 and 7. This research is also supported work funded from the European Research Council (ERC) the European Union’s Horizon 2020 research and innovation programme (grant agreement n◦803193/BEBOP). AJB was supported by STFC grants ST/W000873/1 and UKRI1179. 

The authors would like to attract attention on the help and kind attention of the ESO staff at La Silla and on the dedication of the many technicians and observers from the University of Geneva, to upkeep the telescope and acquire the data that we present here. We would also like to acknowledge that the Euler Swiss Telescope at La Silla is a project funded by the Swiss National Science Foundation (SNSF).

\section*{Data Availability Statement}

All radial velocities and light curves will be made available online.

\renewcommand{\arraystretch}{1.5}
\begin{table*}
\caption{Final fitted parameter values for the joint RV, RM, and photometric fit of EBLM J0021-16. For each fitted parameter, we include the prior values and their distribution that we ran in the code. $\mathcal{N}$ refers to a Normal distribution of a given mean and standard deviation. $\mathcal{U}$ refers to a uniform distribution of given upper and lower bounds}. We also include the timescales for synchronization and circularization as calculated in Sethi et al. (in prep) using Eq. \ref{Synchronization Eq} and \ref{Circularization Eq} respectively. The realignment timescale was calculated using Eq. \ref{Realignment Eq} from \citet{B2016}. The parameters are separated by physical, observable, and RM-specific parameters.   \label{table:MCMC Params}
\begin{tabular}{llll}
\hline
Parameter & Description & Value & Prior Distribution\\
                                                    \hline \hline
\multicolumn{4}{c}{Physical Parameters} \\ \hline
$M_{\rm A}$ ($M_{\odot}$) & Primary Mass & 1.0522 $\pm$ 0.0102 & $\mathcal{N}(1.05, 0.01)$ \\
$M_{\rm B}$ ($M_{\odot}$) & Secondary Mass & 0.1877 $\pm$ 0.0011 & Derived \\
$q$ & Mass Ratio & 0.17843 $\pm$ 0.00062 & $\rm \mathcal{N}(0.19, 1)$ \\
$R_{\rm A}$ ($R_{\odot}$) & Primary Radius & 1.4388 $\pm$ 0.0095 & $\rm \mathcal{N}(1.5, 0.01)$ \\
$R_{\rm B}$ ($R_{\odot}$) & Secondary Radius & 0.2214 $\pm$ 0.0019 & Derived \\
$R_{\rm B}/R_{\rm A}$ & Radius Ratio & 0.15389 $\pm$ 0.00069 & $\rm \mathcal{U}(0.02, 1)$ \\
$S_{\rm B} / S_{\rm A}$ & Surface Brightness Ratio & 0.0432 $\pm$ 0.0013 & $\rm \mathcal{N}(0.14, 1)$\\
$\rho$ (g/cm$^3$) & Stellar Density & 0.4969 $\pm$ 0.0084 & Derived\\
$T_{\rm A}$ (K) & Primary Effective Temperature & 5800 $\pm$ 155 & $\rm \mathcal{N}(5800, 155)$ \\
$T_{\rm B}$ (K) & Secondary Effective Temperature & 2933 $\pm$ 49 & Derived\\
\hline \multicolumn{4}{c}{Orbital Parameters} \\ \hline
$T_{\rm 0}$ (Days) & Time of Eclipse & -1551.28184 $\pm$ 0.00019 & $\rm \mathcal{N}(-1552, 10)$ \\
$P_{\rm orb}$ (Days) & Orbital Period & 5.96727088 $\pm$ 0.00000034 & $\rm \mathcal{N}(5.97, 0.1)$\\
$b$ & Impact Parameter & 0.5961 $\pm$ 0.0089 & $\rm \mathcal{U}(0, 1.16)$ \\
$e$ & Eccentricity & 0.000534 $\pm$ 0.000095 & Derived \\
$\omega$ $(^{\circ})$ & Argument of Periastron & -6 $\pm$ 16 & Derived \\
$\gamma$ (km/s) & Systemic Velocity & 4.5730 $\pm$ 0.0027 & $\rm \mathcal{N}(4.27, 1)$ \\
$\dot{\gamma}$ (m/s/year) & RV Linear Trend & -28.71 $\pm$ 0.80 & $\rm \mathcal{N}(0, 36500)$ \\
$I_{\rm bin}$ $(^{\circ})$ & Binary Inclination & 86.694 $\pm$ 0.066 & Derived \\
$a$ ($R_{\odot}$) & Semi-major Axis & 14.873 $\pm$ 0.045 & Derived \\
$a$ (AU) & Semi-major Axis & 0.06917  $\pm$ 0.00021 & Derived \\

\hline \multicolumn{4}{c}{RM Parameters} \\ \hline
$v_{\rm rot} \sin I_{\rm \star}$ (km/s) & Projected Rotational Velocity & 8.76 $\pm$ 0.17 & $\rm \mathcal{N}(3, 0.5)$ \\
$I_{\rm \star}$ $(^{\circ})$ & Stellar Inclination & 57.9 $\pm$ 2.1 & Derived \\
$|\lambda|$ $(^{\circ})$ & Sky-Projected Obliquity & 2.0 $\pm$ 1.1 & $\rm \mathcal{U}(-180, 180)$ \\
$\psi$ $(^{\circ})$ & True Obliquity & 28.9 $\pm$ 2.1 & Derived \\
$\zeta$ (km/s) & Macroturbulent Velocity & 4.84 $\pm$ 0.96 & $\rm \mathcal{U}(0.5, 6.5)$ \\
$\beta$ (km/s) & Thermal Velocity & 0.64 $\pm$ 0.34 & $\rm \mathcal{U}(0, 1.2)$ \\
$\gamma$ (km/s) & Lorentzian Velocity & 1.32 $\pm$ 0.63 & $\rm \mathcal{U}(0.5, 4.5)$ \\
$P_{\rm rot}$ (days) & Rotational Period & 7.043 $\pm$ 0.060 & \citet{Sethi2024} \\
\hline \multicolumn{4}{c}{Theoretical Timescales} \\ \hline
$\tau_\psi$ (Myr) & Realignment Timescale & 10 & Eq.~\ref{Realignment Eq} \\
$\tau_{\rm sync}$ (Myr) & Synchronization Timescale & 3 & Eq.~\ref{Synchronization Eq} \\
$\tau_{\rm circ}$ (Myr) & Circularization Timescale & 84 & Eq.~\ref{Circularization Eq} \\

\hline
\end{tabular}
\end{table*}

\bibliographystyle{mnras}
\bibliography{EBLMIX} 


\renewcommand{\arraystretch}{1.5}

\newpage

\bsp	
\label{lastpage}
\end{document}